\documentclass[12pt,journal,onecolumn]{IEEEtran}
\usepackage{mathrsfs}

\usepackage{graphicx,float}
\usepackage{color}
\usepackage{amsmath,amsfonts,amssymb}
\usepackage{cite}
\usepackage{alltt}

\usepackage{times} 
\usepackage{epsfig}
\usepackage{pgf,tikz}
\usetikzlibrary{arrows}
\usepackage{algpseudocode}
\usepackage{pgf,tikz}
\usepackage{mathrsfs}
\usepackage{bm}

\begin{document}

\newtheorem{theorem}{Theorem}
\newtheorem{lemma}{Lemma}
\newtheorem{proposition}{Proposition}
\newtheorem{definition}{Definition}
\newtheorem{remark}{Remark}
\newtheorem{corollary}{Corollary}

\title{\LARGE Sparse System Identification for Stochastic Feedback Control Systems
\thanks{Email: \texttt{wxzhao@amss.ac.cn; gyin@wayne.edu; er-wei-bai@uiowa.edu}.}
\thanks{The research of Wenxiao Zhao was supported by National Key Research and Development Program of China (2016YFB0901900) and the National Nature Science Foundation of China under Grants with No. 61822312 and 61573345. The research of
George Yin was supported in part by the Air Force Office of Scientific Research under grant FA9550-18-1-0268. The research of
Er-Wei Bai was supported by the U.S. National Science Foundation.}
}

\author{\normalsize Wenxiao Zhao$^{*}$,~~George Yin$^{**}$,~~Er-Wei Bai$^{\ddag}$\\
\small
$^*$Key Laboratory of Systems and Control, Academy of Mathematics and Systems Science, Chinese
Academy of Sciences, Beijing 100190, China.\\
School of Mathematical Sciences, University of Chinese Academy of Sciences, Beijing 100049, China.\\
$^{**}$Department of Mathematics, Wayne State University, Detroit, MI 48202, USA.\\
$^{\ddag}$Department of Electrical and Computer Engineering, University of Iowa, Iowa City, IA 52242, USA.
}

\date{}
\maketitle

\vspace{-1cm}

\begin{abstract}
Focusing on
identification,
this paper develops techniques to reconstruct
 zero and nonzero elements of a sparse parameter vector $\theta$ of a stochastic dynamic system under feedback control, for which the current input may depend on the past inputs and outputs, system noises as well as exogenous dithers. First, a sparse parameter identification algorithm is introduced based on $L_2$ norm with $L_1$ regularization, where the adaptive weights are adopted in the optimization variables of $L_1$ term. Second, estimates generated by the algorithm are shown to have both set and parameter convergence. That is, sets of the zero and nonzero elements in the parameter $\theta$ can be correctly identified with probability one using a finite number of observations, and estimates of the nonzero elements converge to the true values almost surely. Third, it is shown
 that the results
 are applicable to a large number of applications, including variable selection, open-loop identification, and closed-loop control of stochastic systems. Finally, numerical examples are given to
 support the theoretical analysis.
\end{abstract}

\IEEEpeerreviewmaketitle

\section{Introduction}

Sparsity of a parameter vector in stochastic dynamic systems and precise reconstruction of its zero and nonzero elements
appear
in many areas including systems and control \cite{BaiLi}\cite{Bonin}\cite{Chiuso2014}\cite{Morici}\cite{Toth}\cite{Xiong}, signal processing \cite{Candes}\cite{MaoBillings}\cite{Roweis}, statistics \cite{Knight}\cite{ZhaoYu}\cite{Zou}, and machine learning \cite{Debruyne}\cite{Ojeda}. From a systematic viewpoint it provides a way to discover a parsimonious model that leads to a more reliable prediction model. Classical system identification theory has been a well developed field and achieved great success in both theoretical research and practical applications \cite{ChenGuo}\cite{Ljung87}\cite{Pintelon}. It usually characterizes the identification error between the estimates and the unknown parameters using different criteria
such as randomness of noises, frequency domain sample data, and uncertainty bound of system, etc., so that consistency, convergence rate, and asymptotical normality of estimates can be established as the number of data points goes to infinity. However, these theory and methods are ill suited for sparse identification if the dimension is high.

Over the last few years considerable progress has been made in the precise reconstruction of the zero and nonzero elements in an unknown sparse parameter vector of a stochastic dynamics system based on its input and output observations, for example, the compressed sensing (CS) based identification methods \cite{Perepu}\cite{Toth} and the corresponding adaptive/online algorithms \cite{Chen}\cite{Kalouptsidis}\cite{Kopsinis}, the variable selection algorithms \cite{Cox}\cite{Knight}\cite{ZhaoYu}, etc. The basic idea of CS theory is to obtain a sparsest estimate of the parameter vector by minimizing the $L_0$ norm, i.e., the number of nonzero elements, with $L_2$ constraints \cite{Candes}\cite{Donoho}. The computational complexity for solving $L_0$ minimization problem is NP-hard in general,
 which
 leads to replacing $L_0$ norm with $L_1$ norm, which can be effectively solved by convex optimization techniques. Combining this idea and the dynamics of systems, in \cite{Perepu}\cite{Toth}\cite{Xiong} the CS method is applied to the parameter  estimation of linear systems and in \cite{Chen}\cite{Kalouptsidis}\cite{Kopsinis} the adaptive algorithms such as least mean square (LMS), Kalman filtering (KF), Expectation Maximization (EM), and projection operator are introduced. The variable selection problem aims to find the true but unknown contributing variables of a system among many alternative ones. This often leads to inferring the corresponding parameters being zero or not and estimate the values of the nonzero ones. Classical variable selection algorithms includes MDS \cite{Cox}, LASSO and its variants \cite{Knight}\cite{ZhaoYu}\cite{Zou}, the correlation coefficient method
\cite{WeiBillings2007}, mutual information method \cite{WeiBillings2008}, Bayesian
method \cite{Chiuso2014}, and kernel-based method \cite{Pillonetto2013}, etc. The LASSO-type estimator is usually formulated as the $L_2$ modeling error with $L_1$ regularization, which is also called the basis pursuit in CS literature \cite{Donoho}. Related methods also include those from machine learning perspective \cite{Debruyne}\cite{Ojeda}.

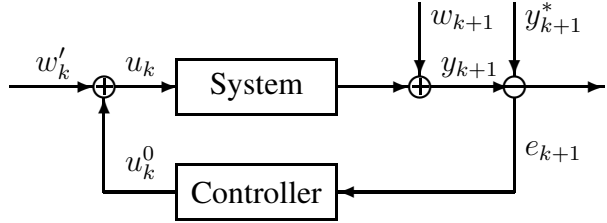
\begin{figure}[hbp]
\centering \setlength{\unitlength}{0.7cm}
\begin{picture}(12,4)
\thicklines
\put(0,2.5){\vector(1,0){1.6}} \put(0.5,2.8){$w'_k$} \put(2.2,2.8){$u_k$}
\put(1.8,2.5){\circle{0.4}}\put(1.665,2.5){\line(1,0){0.3}} \put(1.815,2.35){\line(0,1){0.3}}
\put(2,2.5){\vector(1,0){1.2}}
\put(3.2,2){\framebox(3,1)[]{System}}
\put(6.2,2.5){\vector(1,0){1.4}} 
\put(7.8,2.5){\circle{0.4}} \put(7.665,2.5){\line(1,0){0.3}} \put(7.815,2.35){\line(0,1){0.3}}
\put(7.8,4.1){\vector(0,-1){1.4}}\put(8,3.7){$w_{k+1}$}
\put(8,2.5){\vector(1,0){1.5}} \put(8.2,2.8){$y_{k+1}$}
\put(9.6,2.5){\circle{0.4}} \put(9.46,2.5){\line(1,0){0.3}}
\put(9.6,4.1){\vector(0,-1){1.4}}\put(9.8,3.7){$y^*_{k+1}$}
\put(9.8,2.5){\vector(1,0){1.5}} 
\put(9.6,2.3){\line(0,-1){1.8}}\put(9.8,1.2){$e_{k+1}$}
\put(9.6,0.5){\vector(-1,0){3.35}}
\put(3.2,0){\framebox(3,1)[]{Controller}}
\put(3.15,0.5){\line(-1,0){1.35}}
\put(1.8,0.5){\vector(0,1){1.8}}\put(2.2,0.9){$u^0_{k}$}
\end{picture}
\caption{Flow Diagram of Stochastic Feedback Control System}\label{Fig1}
\end{figure}

In all of the above literature, the random signals are usually assumed to be independent and identically distributed (i.i.d.) or with {\em a prior} knowledge on the sample probability distribution. To the authors' best knowledge, there is no consistent result for sparse parameter identification of stochastic systems with feedback control, which plays a central role in systems and control field. For a general form of feedback control, inputs at the current time depend on the past inputs and outputs, system noises, and possibly exogenous dithers; see, e.g., Fig.\ref{Fig1}, where $y_{k+1}$ is the system output, $w_{k+1}$ is the system noise, $y^*_{k+1}$ is the tracking signal or regulation signal, $w'_{k}$ denotes some exogenous input, and $u_{k}$ is the feedback control so that for the closed-loop system the error $e_{k+1}=|y_{k+1}-y^*_{k+1}|$ is minimized in some sense. A problem closely related to sparse parameter identification is the order estimation, which estimates the maximal time index for characterizing the dependence between the current output and the past inputs and outputs. It has been extensively studied in the literature, for example, the Akaike's information criterion (AIC) for stationary time series \cite{Akaike} and the control information criterion (CIC) for linear feedback control systems \cite{ChenGuo}. Compared with the order estimation, the sparse parameter identification in fact goes further: once the zero and nonzero elements being correctly identified, estimates of the system order follow directly.

In this paper, we
consider sparse parameter identification of a stochastic linear system with feedback control as illustrated in Fig. \ref{Fig1}. Here the framework is not confined to specific types of feedback control law, such as PID control, adaptive regulation control, or model reference control, etc., but in a general form. That is, the control input $u_{k+1}$ belongs to the $\sigma$-algebra $\mathcal{F}_{k+1}$ generated by the past system inputs, outputs, noises, and exogenous dithers.

Our contributions of the
paper are as follow.
First, a sparse parameter identification algorithm is proposed, which is based on $L_2$ estimation error with $L_1$ regularization. The key difference between the proposed algorithm and those in CS and LASSO framework lies in that weighting coefficients generated from the data of closed-loop systems are introduced to each of the optimization variables in the $L_1$ term. Second, under suitable
conditions we prove that estimates generated from the algorithms have both set and parameter convergence, that is, sets of the zero and nonzero elements in the unknown sparse parameter vector can be identified with probability one with a finite number of observations, which is different from the asymptotical theory in the classical system identification literature, and furthermore, estimates of the nonzero ones converge to the true values almost surely. Third, we will show that the usual persistent excitation (PE) condition for system identification and irrepresentable conditions for consistency of LASSO, are not required, which relies on the {\em a prior} information on the sets of the zero and nonzero elements in the parameter vector as well as the data matrix, see, e.g., Table \ref{tab2} in Section III. We also apply the proposed algorithm to identification of open-loop Hammerstein systems and closed-loop linear stochastic systems with adaptive regulation control, both with sparse parameters. The Hammerstein system, consisting of a static nonlinear function followed by a linear subsystem, is a good approximation to real systems in chemical engineering, biological cybernetics, etc., and has been widely studied in identification and engineering practice \cite{Sjoberg}. The adaptive regulation control of linear systems, or the celebrated self-tuning regulator, received much attention during the seventies and eighties of the last century and has been successfully applied in practice \cite{Astrom}\cite{Guo}. By applying the proposed algorithm to the identification of the two systems, both set convergence and parameter convergence are obtained and strong consistency of estimates is established.

The rest of the paper is organized as follow.
Problem formulation and algorithm design are given in Section II and the theoretical results are established in Section III. In Section IV, we compare the technical conditions in this paper with the regular and irrepresentable conditions for LASSO and we also apply the algorithm to the parameter estimation of Hammerstein system and linear stochastic system with adaptive regulation control. In Section V, we 
present
simulation examples to illustrate performance of the algorithm and in Section VI, we given some concluding remarks.


{\em Notation.} Let $(\Omega,\mathscr{F},\mathbb{P})$ be the probability space,
$\omega$ be an element in $\Omega$, and $\mathbb{E}(\cdot)$ be the expectation
operator. By $\|\cdot\|_0$, $\|\cdot\|_1$, and $\|\cdot\|_2$ we denote the $0$-norm, $1$-norm, and $2$-norm of vectors or matrices, respectively. In this paper the $2$-norm is also denoted by $\|\cdot\|$ for simplicity and the Frobenous norm is dented by $\|\cdot\|_F$. For
two positive sequences $\{a_k\}_{k\geq1}$ and $\{b_k\}_{k\geq1}$, by $a_k=O(b_k)$ we
mean $a_k\leq cb_k,~k\geq1$ for some $c>0$ while $a_k=o(b_k)$ means $a_k/b_k\to0$ as
$k\to\infty$. By $\mathrm{sgn}(x)$ we denote the sign function, i.e.,
$\mathrm{sgn}(x)=1~\mathrm{if}~x\geq0$ and$~\mathrm{sgn}(x)=-1~\mathrm{if}~x<0$. The maximal and minimal eigenvalues of a symmetric matrix $M$ are denoted by $\lambda_{\max}\{M\}$ and $\lambda_{\min}\{M\}$, respectively.

\section{Problem Formulation and Sparse Identification Algorithm}

Consider the parameter identification of the following stochastic system,
\begin{align}
y_{k+1}=\theta^T\varphi_k+w_{k+1},~~k\geq0,\label{Eq1}
\end{align}
where $\theta$ is the unknown $r$-dimensional parameter vector, $\varphi_k\in \mathbb{R}^r$, with $r$ being known consisting possibly current and past inputs and outputs, is the regressor vector, $y_{k+1}$ and $w_{k+1}$ are the system output and noise, respectively.

Denote the family of $\sigma$ algebras
$\{\mathcal{F}_k\}$ by
\begin{align}
\mathcal{F}_k&\triangleq  \sigma\{y_k,\dots,y_0,u_{k-1},\dots,u_0,w_k,\dots,w_0,w'_k,\dots,w'_0\},~~k\geq1\label{Eq2}
\end{align}
where $\{w'_k\}$ is the sequence of exogenous input signals to the system. See, e.g., Fig. \ref{Fig1}.
Moreover,
denote the parameter vector $\theta$ and the index set of its zero elements by
\begin{align}
\theta&\triangleq[\theta(1)\dots \theta(r)]^T\label{Eq3}\\
A^*&\triangleq\{j=1,\dots,r~\big|~\theta(j)=0\}.\label{Eq4}
\end{align}
By assuming that the regressor $\varphi_k$ is $\mathcal{F}_k$-measurable for each $k\geq1$, the problem is to infer the set $A^*$ and to estimate the unknown but nonzero elements in $\theta$ based on the system observations $\{\varphi_k,y_{k+1}\}_{k=1}^N$.

\begin{remark}\label{Remark1}
If $\varphi_k=[y_k,\dots,y_{k+1-p},u_{k},\cdots,u_{k+1-q}]^T$ and $\theta=[a_1,\dots,a_p,b_1,\cdots,b_q]^T$, then system (\ref{Eq1}) falls into the classical ARX system. In addition, it can also include the parameterized nonlinear systems such as Hammerstein system \cite{Zhao2010}, nonlinear ARX system \cite{Sjoberg}, etc.
\end{remark}

We now introduce the sparse identification algorithm for $\theta$, which consists of two steps, first to estimate $\theta$ with the least squares (LS) algorithm and then, based on the estimates in the first step to construct a convex optimization problem
and to further identify
the sets of zero and nonzero elements in $\theta$.

Denote the maximal and minimal eigenvalues of $\sum\limits_{k=1}^N\varphi_k\varphi_k^T$ by $\lambda_{\max}(N)$ and $\lambda_{\min}(N)$, respectively.

\noindent
\textbf{Algorithm}
\begin{alltt}
Step 1. Based on \(\{\varphi\sb{k},y\sb{k+1}\}\sb{k=1}\sp{N}\), compute the LS estimate of \(\theta\)
\end{alltt}
\begin{align}
\theta_{N+1}=\left(\sum\limits_{k=1}^N\varphi_k\varphi_k^T\right)^{-1}
\left(\sum\limits_{k=1}^N\varphi_ky_{k+1}\right).\label{Eq5}
\end{align}
\begin{alltt}
Denote
\end{alltt}
\begin{align}
\theta_{N+1}\triangleq[\theta_{N+1}(1),\cdots,\theta_{N+1}(r)]^T,\label{Eq6}
\end{align}
\begin{alltt}
and further define
\end{alltt}
\begin{align}
\widehat{\theta}_{N+1}(j)\triangleq \theta_{N+1}(j)+\mathrm{sgn}\big(\theta_{N+1}(j)\big)\sqrt{\frac{\log \lambda_{\max}(N)}{\lambda_{\min}(N)}}.\label{Eq7}
\end{align}

\begin{alltt}
Step 2. Construct the convex optimization algorithm
\end{alltt}
\begin{align}
J_{N+1}(\beta)\triangleq&\sum\limits_{k=1}^N(y_{k+1}-\beta^T\varphi_k)^2+\lambda_N\sum\limits_{l=1}^r \frac{1}{|\widehat{\theta}_{N+1}(l)|}|\beta(l)|\label{Eq8}
\end{align}
\begin{alltt}
for some \(\lambda\sb{N}>0\) and
\end{alltt}
\begin{align}
\beta_{N+1}=&[\beta_{N+1}(1)\dots \beta_{N+1}(r)]^T\triangleq \mathop{\mathrm{argmin}}\limits_{\beta}J_{N+1}(\beta)\label{Eq9}\\
A^*_{N+1}\triangleq&\{j=1,\dots,r~|~\beta_{N+1}(j)=0\}.\label{Eq10}
\end{align}

\begin{remark}\label{Remark2}
The set $A^*_{N+1}$ generated from the convex optimization problem
(\ref{Eq8}) serves as the estimate for set $A^*$ of the zero elements in $\theta$. The coefficient sequence $\{\lambda_N\}$ in (\ref{Eq8}) is chosen as a positive sequence tending to infinity, which will be specified later. Note
that $\widehat{\theta}_{N+1}(l)$ appears in the denominator of algorithm (\ref{Eq8}). Thus if $\widehat{\theta}_{N+1}(l)\to0$ as $N\to\infty$ for some $l=1,\dots,r$ and hence $\frac{1}{|\widehat{\theta}_{N+1}(l)|}\to\infty$, then the corresponding minimizer $\beta_{N+1}(l)$ should be exactly zero. This explains why algorithm (\ref{Eq8}) generates sparse solution. The modification of the LS estimates given by (\ref{Eq7}) is to prevent them from zero since otherwise algorithm (\ref{Eq8}) would be insignificant.
\end{remark}

\begin{remark}\label{Remark3}
In the literature, the basic pursuit refers to solve the following convex optimization problem:
\begin{align}
\min\limits_{x}\frac{1}{2}\|y-Mx\|^2_2+\lambda\|x\|_1.\label{Eq11}
\end{align}
In
the CS theory, LASSO method for variable selection and algorithm (\ref{Eq8}) all fall into this category \cite{Gill}.
 Note that in \cite{Zou} a modified LASSO-type estimator with weights adopted in the $L_1$ regularization term is also introduced. The essential difference between algorithm (\ref{Eq8}) and that in \cite{Zou} lies in the fact that in this paper the data sequence $\{\varphi_k,y_{k+1}\}_{k\geq1}$ admits feedback control while the conditions in \cite{Zou} do not include this case.
\end{remark}

Next we introduce assumptions to be used for the theoretical analysis.

\begin{itemize}
\item[A1)] The noise
$\{w_k,\mathcal{F}_k\}_{k\geq1}$ is a martingale difference sequence, i.e., $\mathbb{E}[w_{k+1}|\mathcal{F}_k]=0,~k\geq1$, and there exists some $\gamma>2$ such that $\sup\limits_k\mathbb{E}\big[|w_{k+1}|^{\gamma}|\mathcal{F}_k\big]
<\infty~\mathrm{a.s.}$
\item[A2)] For each $k\geq1$, $\varphi_k$ is $\mathcal{F}_k$-measurable.
\item[A3)] For the maximal and minimal eigenvalues of $\sum\limits_{k=1}^N\varphi_k\varphi_k^T$, it holds that \begin{align}
    \frac{\lambda_{\max}(N)}{\lambda_{\min}(N)}\sqrt{\frac{\log \lambda_{\max}(N)}{\lambda_{\min}(N)}}
    \mathop{\longrightarrow}\limits_{N\to\infty}0~~\mathrm{a.s.}\label{Eq12}
    \end{align}
\item[A4)] $\{\lambda_N\}$ in algorithm (\ref{Eq8}) is a positive sequence such that
\begin{align}
\lambda_N=o\left(\lambda_{\min}(N)\right),~~\lambda_{\max}(N)\sqrt{\frac{\log \lambda_{\max}(N)}{\lambda_{\min}(N)}}=o\left(\lambda_N\right)~~\mathrm{a.s.}\label{Eq13}
\end{align}
\end{itemize}

\begin{remark}\label{Remark4}
We can directly verify that if $\{w_k\}$ is a sequence of i.i.d. Gaussian variables, then A1) holds true for any fixed $\gamma>2$ and A2) is satisfied for a large number of systems such as PID control, adaptive regulation control and model reference control etc. Assumptions A3) and A4) is
a weak condition  compared with
the traditional persistent excitation condition.
In fact,
$\frac{1}{N}\sum\limits_{k=1}^N\varphi_k\varphi_k^T$ tending to a positive definite matrix is not required.
\end{remark}

\section{Theoretical Properties of Sparse Identification Algorithm}

\subsection{Set and Parameter Convergence of Estimates}

Assume that there are $d$ nonzero elements in vector $\theta$. Without losing generality, we assume $\theta=[\theta(1)\dots\theta(d)~\theta(d+1)\dots\theta(r)]^T$ and $\theta(i)\neq0,~i=1,\dots,d,~\theta(j)=0,~j=d+1,\dots,r$.
For the estimate $\beta_{N+1}$ generated by algorithms (\ref{Eq5})--(\ref{Eq9}), we have the following result.

\begin{theorem}\label{Thm1}
Assume that A1)-A4) holds. Then there exists an $\omega$-set $\Omega_0$ with $\mathbb{P}\{\Omega_0\}=1$ such that for any $\omega\in\Omega_0$, there exists an integer $N_0(\omega)$ such that
\begin{align}
\beta_{N+1}(d+1)=\cdots=\beta_{N+1}(r)=0,~~N\geq N_0(\omega)\label{Eq16}
\end{align}
and
\begin{align}
\beta_{N+1}(i)\mathop{\longrightarrow}
\limits_{N\to\infty}\theta(i),~~i=1,\cdots,d.~~\mathrm{a.s.}\label{Eq17}
\end{align}
\end{theorem}

Theorem \ref{Thm1} shows that the index set of the zero elements in $\theta$ can be correctly identified with a finite number of observations, i.e., $A_{N+1}^*=A^*$ for all $N$ large enough and estimates for the nonzero elements converge to the true values with probability one. Noticing that the criterion function (\ref{Eq8}) is convex, thus a variety of efficient numerical methods can be applied to obtain the estimate $\beta_{N+1}$. Next we prove Theorem \ref{Thm1}. Before giving the proof, we first state two classical results in stochastic adaptive control.

\begin{lemma}\label{Lem1}(\cite{LaiWei})
Assume that A1) and A2) hold. Then as $N\to\infty$,
\begin{align}
\left\|\left(\sum\limits_{k=1}^N\varphi_k\varphi_k^T\right)^{-\frac{1}{2}} \sum\limits_{k=1}^N\varphi_kw_{k+1}\right\|
=O\left(\sqrt{\log \lambda_{\max}(N)}\right).\label{Eq14}
\end{align}
\end{lemma}

\begin{lemma}\label{Lem2}(\cite{ChenGuo})
Assume that A1) and A2) hold. Then as $N\to\infty$, the estimation error of the LS algorithm is bounded by
\begin{align}
\left\|\theta_{N+1}-\theta\right\|^2=O\left(\frac{\log\lambda_{\max}(N)}{\lambda_{\min}(N)}\right) ~~\mathrm{a.s.}\label{Eq15}
\end{align}
\end{lemma}

{\em Proof of Theorem \ref{Thm1}:} Noting
that A3) and A4) hold almost surely,
there exists $\Omega_0$ with $\mathbb{P}\{\Omega_0\}=1$ such that A3) and A4) hold for any $\omega\in\Omega_0$. In the following, we will consider the estimate sequence $\{\beta_{N+1}\}$ on a fixed sample path $\omega\in\Omega_0$.

Denote the estimate $\beta_{N+1}$ by
\begin{align}
\beta_{N+1}=\theta+\mu_{N+1}.\label{Eq18}
\end{align}
For (\ref{Eq16}) and (\ref{Eq17}), it suffices to prove that there exists $N_0$ large enough such that
\begin{align}
&\mu_{N+1}(d+1)=\cdots=\mu_{N+1}(r)=0,~~N\geq N_0,\label{Eq19}
\end{align}
and
\begin{align}
&\mu_{N+1}(l)\mathop{\longrightarrow}\limits_{N\to\infty}0,~~l=1,\dots,d.\label{Eq20}
\end{align}

The proof can be divided into two steps. First
prove $\mu_{N+1}(l)\mathop{\longrightarrow}\limits_{N\to\infty}0,~~l=1,\dots,r$, and then show
$\mu_{N+1}(d+1)=\cdots=\mu_{N+1}(r)=0$ for
$N$ large enough.
Denote by $\{\mu_{N_n+1}\}_{n\geq1}$ the subsequence of $\{\mu_{N+1}\}_{N\geq1}$ such that $\|\mu_{N_n+1}\|>0$ for each $n\geq1$.
By noting
that $\beta_{N+1}=\theta+\mu_{N+1}$ is the minimizer of $J_{N+1}(\beta)$, it follows that
\begin{align}
J_{N+1}(\theta+\mu_{N+1})-J_{N+1}(\theta)\leq0.\label{Eq21}
\end{align}

By (\ref{Eq1}),
(\ref{Eq8}), and noting
$\theta(d+1)=\cdots=\theta(r)=0$, direct calculation leads to
\begin{align}
\nonumber&J_{N+1}(\theta+\mu_{N+1})\\
\nonumber=&\sum\limits_{k=1}^N\left(y_{k+1}-(\theta+\mu_{N+1})^T\varphi_k\right)^2+\lambda_N\sum\limits_{l=1}^r \frac{1}{|\widehat{\theta}_{N+1}(l)|}\left|\theta(l)+\mu_{N+1}(l)\right|\\
\nonumber=&\sum\limits_{k=1}^N\left(w_{k+1}-\mu_{N+1}^T\varphi_k\right)^2+\lambda_N\sum\limits_{l=1}^d \frac{1}{|\widehat{\theta}_{N+1}(l)|}\left|\theta(l)+\mu_{N+1}(l)\right|+\lambda_N\sum\limits_{l=d+1}^r \frac{1}{|\widehat{\theta}_{N+1}(l)|}\left|\mu_{N+1}(l)\right|\\
\nonumber=&\sum\limits_{k=1}^Nw_{k+1}^2-2\mu^T_{N+1}\sum\limits_{k=1}^N\varphi_kw_{k+1} +\mu^T_{N+1}\sum\limits_{k=1}^N\varphi_k\varphi_k^T\mu_{N+1}\\
&+ \lambda_N\sum\limits_{l=1}^d \frac{1}{|\widehat{\theta}_{N+1}(l)|}\left|\theta(l)+\mu_{N+1}(l)\right|+\lambda_N\sum\limits_{l=d+1}^r \frac{1}{|\widehat{\theta}_{N+1}(l)|}\left|\mu_{N+1}(l)\right|\label{Eq22}
\end{align}
and
\begin{align}
\nonumber&J_{N+1}(\theta)\\
\nonumber=&\sum\limits_{k=1}^N\left(y_{k+1}-\theta^T\varphi_k\right)^2+\lambda_N\sum\limits_{l=1}^d \frac{1}{|\widehat{\theta}_{N+1}(l)|}\left|\theta(l)\right|\\
=&\sum\limits_{k=1}^Nw_{k+1}^2+\lambda_N\sum\limits_{l=1}^d \frac{1}{|\widehat{\theta}_{N+1}(l)|}\left|\theta(l)\right|.\label{Eq23}
\end{align}

From (\ref{Eq22}) and (\ref{Eq23}), we have
\begin{align}
\nonumber&J_{N+1}(\theta+\mu_{N+1})-J_{N+1}(\theta)\\
\nonumber=&\mu^T_{N+1}\sum\limits_{k=1}^N\varphi_k\varphi_k^T\mu_{N+1} -2\mu^T_{N+1}\sum\limits_{k=1}^N\varphi_kw_{k+1}\\
\nonumber&+ \lambda_N\sum\limits_{l=1}^d \frac{1}{|\widehat{\theta}_{N+1}(l)|}\left(\left|\theta(l)+\mu_{N+1}(l)\right|-|\theta(l)|\right)\\ &+\lambda_N\sum\limits_{l=d+1}^r \frac{1}{|\widehat{\theta}_{N+1}(l)|}\left|\mu_{N+1}(l)\right|.\label{Eq24}
\end{align}
Define
\begin{align}
&M_{N+1}\triangleq \mu^T_{N+1}\sum\limits_{k=1}^N\varphi_k\varphi_k^T\mu_{N+1} -2\mu^T_{N+1}\sum\limits_{k=1}^N\varphi_kw_{k+1},\label{Eq25}\\
&\alpha_{N+1}\triangleq \left(\sum\limits_{k=1}^N\varphi_k\varphi_k^T\right)^{\frac{1}{2}}\mu_{N+1}.\label{Eq26}
\end{align}

Noting
that $\|\mu_{N_n+1}\|>0$, it can be directly verified that
\begin{align}
M_{N_n+1}=\alpha_{N_n+1}^T \left(I-2\left(\sum\limits_{k=1}^{N_n}\varphi_k\varphi_k^T\right)^{-\frac{1}{2}} \left(\sum\limits_{k=1}^{N_n}\varphi_kw_{k+1}\right)\frac{\mu_{N_n+1}^T}{\|\mu_{N_n+1}\|^2} \left(\sum\limits_{k=1}^{N_n}\varphi_k\varphi_k^T\right)^{-\frac{1}{2}}\right)\alpha_{N_n+1}.\label{Eq27}
\end{align}
By Lemma \ref{Lem1}, we have
\begin{align}
\left\|\left(\sum\limits_{k=1}^{N_n}\varphi_k\varphi_k^T\right)^{-\frac{1}{2}} \left(\sum\limits_{k=1}^{N_n}\varphi_kw_{k+1}\right)\right\|=O\left(\sqrt{\log \lambda_{\max}(N_n)}\right).\label{Eq28}
\end{align}

By the definition of matrix $2$-norm,
\begin{align}
\left\|\left(\sum\limits_{k=1}^{N_n}\varphi_k\varphi_k^T\right)^{-\frac{1}{2}}\right\|
=\lambda^{\frac{1}{2}}_{\max} \left\{\left(\sum\limits_{k=1}^{N_n}\varphi_k\varphi_k^T\right)^{-1}\right\}
=\lambda^{-\frac{1}{2}}_{\min}(N_n).\label{Eq29}
\end{align}
From (\ref{Eq28}) and (\ref{Eq29}), we obtain
\begin{align}
\nonumber&\left|\alpha_{N_n+1}^T \left(\sum\limits_{k=1}^{N_n}\varphi_k\varphi_k^T\right)^{-\frac{1}{2}} \left(\sum\limits_{k=1}^{N_n}\varphi_kw_{k+1}\right)\frac{\mu_{N_n+1}^T}{\|\mu_{N_n+1}\|^2} \left(\sum\limits_{k=1}^{N_n}\varphi_k\varphi_k^T\right)^{-\frac{1}{2}}\alpha_{N_n+1}\right|\\
\nonumber\leq&\|\alpha_{N_n+1}\|^2\cdot O\left(\sqrt{\log \lambda_{\max}(N_n)}\right)\cdot
\frac{1}{\|\mu_{N_n+1}\|}\cdot\frac{1}{\sqrt{\lambda_{\min}(N_n)}}\\
=&\|\alpha_{N_n+1}\|^2\frac{1}{\|\mu_{N_n+1}\|} O\left(\sqrt{\frac{\log \lambda_{\max}(N_n)}{\lambda_{\min}(N_n)}}\right),\label{Eq30}
\end{align}
which together with (\ref{Eq27}) yields that
\begin{align}
M_{N_n+1}\geq \|\alpha_{N_n+1}\|^2-c_1\|\alpha_{N_n+1}\|^2\frac{1}{\|\mu_{N_n+1}\|} \sqrt{\frac{\log \lambda_{\max}(N_n)}{\lambda_{\min}(N_n)}}\label{Eq31}
\end{align}
for some $c_1>0$.

By the definition of $\alpha_{N_n+1}$, it follows that
\begin{align}
\lambda_{\min}(N_n)\|\mu_{N_n+1}\|^2\leq \|\alpha_{N_n+1}\|^2\leq \lambda_{\max}(N_n)\|\mu_{N_n+1}\|^2.\label{Eq32}
\end{align}
From (\ref{Eq31}) and (\ref{Eq32}), we have
\begin{align}
M_{N_n+1}\geq \lambda_{\min}(N_n)\|\mu_{N_n+1}\|^2-c_1\lambda_{\max}(N_n)\|\mu_{N_n+1}\| \sqrt{\frac{\log \lambda_{\max}(N_n)}{\lambda_{\min}(N_n)}}.\label{Eq33}
\end{align}

By A3) and Lemma \ref{Eq2}, we know that the limits of $\theta_{N+1}(l)$ and $\widehat{\theta}_{N+1}(l)$, $l=1,\dots,d$ are nonzero and hence
\begin{align}
\nonumber&\left|\lambda_{N_n}\sum\limits_{l=1}^d\frac{1}{|\widehat{\theta}_{N_n+1}(l)|} \left(|\theta(l)+\mu_{N_n+1}(l)|-|\theta(l)|\right)\right|\\
\leq&c_2\lambda_{N_n}\sum\limits_{l=1}^d|\mu_{N_n+1}(l)|\leq c_2\lambda_{N_n}\|\mu_{N_n+1}\|\label{Eq34}
\end{align}
where $c_2>0$ is a constant which may change among different lines and for the second inequality the equivalence of vector norms in finite dimension space is applied.

Combining (\ref{Eq24}), (\ref{Eq33}), and (\ref{Eq34}), we obtain
\begin{align}
\nonumber0\geq&\lambda_{\min}(N_n)\|\mu_{N_n+1}\|^2-c_1\lambda_{\max}(N_n)\|\mu_{N_n+1}\| \sqrt{\frac{\log \lambda_{\max}(N_n)}{\lambda_{\min}(N_n)}}-c_2\lambda_{N_n}\|\mu_{N_n+1}\|\\
=&\lambda_{\min}(N_n)\|\mu_{N_n+1}\|\left(\|\mu_{N_n+1}\| -c_1\frac{\lambda_{\max}(N_n)}{\lambda_{\min}(N_n)}\sqrt{\frac{\log \lambda_{\max}(N_n)}{\lambda_{\min}(N_n)}}
-c_2\frac{\lambda_{N_n}}{\lambda_{\min}(N_n)}\right),\label{Eq35}
\end{align}
which by noting
$\|\mu_{N_n+1}\|>0$ implies
\begin{align}
\|\mu_{N_n+1}\| \leq c_1\frac{\lambda_{\max}(N_n)}{\lambda_{\min}(N_n)}\sqrt{\frac{\log \lambda_{\max}(N_n)}{\lambda_{\min}(N_n)}}
+c_2\frac{\lambda_{N_n}}{\lambda_{\min}(N_n)}.\label{Eq36}
\end{align}

Since $\{\mu_{N_n+1}\}_{n\geq1}$ is the subsequence of $\{\mu_{N+1}\}_{N\geq1}$ with $\|\mu_{N_n+1}\|>0$, we further have
\begin{align}
\|\mu_{N+1}\| \leq c_1\frac{\lambda_{\max}(N)}{\lambda_{\min}(N)}\sqrt{\frac{\log \lambda_{\max}(N)}{\lambda_{\min}(N)}}
+c_2\frac{\lambda_{N}}{\lambda_{\min}(N)},\label{Eq37}
\end{align}
which together with A3) and A4) yields that $\|\mu_{N+1}\|\to0$ as $N\to\infty$ and hence $\beta_{N+1}=\theta+\mu_{N+1}\to\theta$.

Next, we prove that $\mu_{N+1}(d+1)=\cdots=\mu_{N+1}(r)=0$ for all $N$ large enough.
Otherwise, if for some $i_l\in\{d+1,\dots,r\}$ and some subsequence $\{N_m\}_{m\geq1}$ such that $\mu_{N_m+1}(i_l)\neq0,~m\geq1$, then $\|\mu_{N_m+1}\|>0,~m\geq1$.

Denote
$$
\mu_{N_m+1}\triangleq\begin{bmatrix}\mu^{(1)}_{N_m+1}\\\mu^{(2)}_{N_m+1}\end{bmatrix}~~
\mathrm{and}~~
\overline{\mu}_{N_m+1}\triangleq\begin{bmatrix}\mu^{(1)}_{N_m+1}\\0\end{bmatrix}
$$
where $\mu^{(1)}_{N_m+1}\in \mathbb{R}^d$ and $\mu^{(2)}_{N_m+1}\in \mathbb{R}^{r-d}$.
Since $\beta_{N+1}=\theta+\mu_{N+1}$ is the minimizer of $J_{N+1}(\beta)$, it follows that
\begin{align}
J_{N_m+1}(\theta+\mu_{N_m+1})-J_{N_m+1}(\theta+\overline{\mu}_{N_m+1})\leq0.\label{Eq38}
\end{align}
Denote
$$
\sum\limits_{k=1}^N\varphi_k\varphi^T_k\triangleq
\begin{bmatrix}\Phi_N^{(11)} & \Phi_N^{(12)}\\\Phi_N^{(21)} & \Phi_N^{(22)}\end{bmatrix}
~~\mathrm{and}~~
\varphi_k\triangleq\begin{bmatrix}\varphi_k^{(1)}\\\varphi_k^{(2)}\end{bmatrix}
$$
where $\Phi_N^{(11)}\in \mathbb{R}^{d\times d}$, $\varphi_k^{(1)}\in \mathbb{R}^d$, and others are with compatible dimensions.

From (\ref{Eq22}) it follows that for $\mu_{N_m+1}$
\begin{align}
\nonumber&J_{N_m+1}(\theta+\mu_{N_m+1})\\
\nonumber=&\sum\limits_{k=1}^{N_m}w_{k+1}^2-2\mu^T_{N_m+1}\sum\limits_{k=1}^{N_m}\varphi_kw_{k+1} +\mu^T_{N_m+1}\sum\limits_{k=1}^{N_m}\varphi_k\varphi_k^T\mu_{N_m+1}\\
\nonumber&+ \lambda_{N_m}\sum\limits_{l=1}^d \frac{1}{|\widehat{\theta}_{N_m+1}(l)|}\left|\theta(l)+\mu_{N_m+1}(l)\right|+\lambda_{N_m}\sum\limits_{l=d+1}^r \frac{1}{|\widehat{\theta}_{N_m+1}(l)|}\left|\mu_{N_m+1}(l)\right|\\
\nonumber=&\sum\limits_{k=1}^{N_m}w_{k+1}^2-2\mu^{(1)T}_{N_m+1}\sum\limits_{k=1}^{N_m}\varphi^{(1)}_kw_{k+1} -2\mu^{(2)T}_{N_m+1}\sum\limits_{k=1}^{N_m}\varphi^{(2)}_kw_{k+1} \\ \nonumber&+\mu^{(1)T}_{N_m+1}\Phi^{(11)}_{N_m}\mu^{(1)}_{N_m+1} +\mu^{(2)T}_{N_m+1}\Phi^{(21)}_{N_m}\mu^{(1)}_{N_m+1}
+\mu^{(1)T}_{N_m+1}\Phi^{(12)}_{N_m}\mu^{(2)}_{N_m+1}
+\mu^{(2)T}_{N_m+1}\Phi^{(22)}_{N_m}\mu^{(2)}_{N_m+1}\\
&+ \lambda_{N_m}\sum\limits_{l=1}^d \frac{1}{|\widehat{\theta}_{N_m+1}(l)|}\left|\theta(l)+\mu_{N_m+1}(l)\right|+\lambda_{N_m}\sum\limits_{l=d+1}^r \frac{1}{|\widehat{\theta}_{N_m+1}(l)|}\left|\mu_{N_m+1}(l)\right|\label{Eq39}
\end{align}
and for $\overline{\mu}_{N_m+1}$
\begin{align}
\nonumber&J_{N_m+1}(\theta+\overline{\mu}_{N_m+1})\\
\nonumber=&\sum\limits_{k=1}^{N_m}w_{k+1}^2 -2\mu^{(1)T}_{N_m+1}\sum\limits_{k=1}^{N_m}\varphi^{(1)}_kw_{k+1} +\mu^{(1)T}_{N_m+1}\Phi^{(11)}_{N_m}\mu^{(1)}_{N_m+1}
\end{align}
\begin{align}
&+ \lambda_{N_m}\sum\limits_{l=1}^d \frac{1}{|\widehat{\theta}_{N_m+1}(l)|}\left|\theta(l)+\mu_{N_m+1}(l)\right|\label{Eq40}
\end{align}
by noting
that $\overline{\mu}_{N_m+1}=[\mu^{(1)T}_{N_m+1}~0]^T$.

From (\ref{Eq39}) and (\ref{Eq40}), we have
\begin{align}
\nonumber&J_{N_m+1}(\theta+\mu_{N_m+1})-J_{N_m+1}(\theta+\overline{\mu}_{N_m+1})\\
\nonumber=&-2\mu^{(2)T}_{N_m+1}\sum\limits_{k=1}^{N_m}\varphi^{(2)}_kw_{k+1} +\mu^{(2)T}_{N_m+1}\Phi^{(22)}_{N_m}\mu^{(2)}_{N_m+1}+\mu^{(2)T}_{N_m+1}\Phi^{(21)}_{N_m}\mu^{(1)}_{N_m+1}\\
&
+\mu^{(1)T}_{N_m+1}\Phi^{(12)}_{N_m}\mu^{(2)}_{N_m+1}+\lambda_{N_m}\sum\limits_{l=d+1}^r \frac{1}{|\widehat{\theta}_{N_m+1}(l)|}\left|\mu_{N_m+1}(l)\right|.\label{Eq41}
\end{align}

By Lemma \ref{Lem1}, we have the following equalities and inequalities,
\begin{align}
\nonumber&\mu^{(2)T}_{N_m+1}\Phi^{(22)}_{N_m}\mu^{(2)}_{N_m+1} -2\mu^{(2)T}_{N_m+1}\sum\limits_{k=1}^{N_m}\varphi^{(2)}_kw_{k+1}\\
\nonumber=&\mu^{(2)T}_{N_m+1}\Phi^{(22)}_{N_m}\mu^{(2)}_{N_m+1} -2\mu^{(2)T}_{N_m+1}\left(\Phi^{(22)}_{N_m}\right)^{\frac{1}{2}} \left(\Phi^{(22)}_{N_m}\right)^{-\frac{1}{2}}\sum\limits_{k=1}^{N_m}\varphi^{(2)}_kw_{k+1}\\
\nonumber\geq &\lambda_{\min}\{\Phi^{(22)}_{N_m}\}\|\mu^{(2)}_{N_m+1}\|^2 -2\left\|\mu^{(2)}_{N_m+1}\right\|\left\|\left(\Phi^{(22)}_{N_m}\right)^{\frac{1}{2}} \right\|\left\| \left(\Phi^{(22)}_{N_m}\right)^{-\frac{1}{2}}\sum\limits_{k=1}^{N_m}\varphi^{(2)}_kw_{k+1}\right\|\\
\geq& \lambda_{\min}\{\Phi^{(22)}_{N_m}\}\|\mu^{(2)}_{N_m+1}\|^2-c_3\lambda^{\frac{1}{2}}_{\max}\{\Phi^{(22)}_{N_m}\} \left\|\mu^{(2)}_{N_m+1}\right\|
\sqrt{\log \lambda_{\max}\{\Phi^{(22)}_{N_m}\}}.\label{Eq42}
\end{align}

Noting
that $\lambda_{\max}\{\Phi^{(22)}_{N_m}\}\leq \lambda_{\max}(N_m)$ and $\lambda_{\min}\{\Phi^{(22)}_{N_m}\}\geq \lambda_{\min}(N_m)$, from (\ref{Eq42}) we obtain
\begin{align}
\nonumber&\mu^{(2)T}_{N_m+1}\Phi^{(22)}_{N_m}\mu^{(2)}_{N_m+1} -2\mu^{(2)T}_{N_m+1}\sum\limits_{k=1}^{N_m}\varphi^{(2)}_kw_{k+1}\\
\nonumber\geq& \lambda_{\min}(N_m)\|\mu^{(2)}_{N_m+1}\|^2-c_3\lambda^{\frac{1}{2}}_{\max}(N_m) \left\|\mu^{(2)}_{N_m+1}\right\|
\sqrt{\log \lambda_{\max}\{N_m\}}\\
\geq& \lambda_{\min}(N_m)\|\mu^{(2)}_{N_m+1}\|^2-c_3\lambda_{\max}(N_m) \left\|\mu^{(2)}_{N_m+1}\right\|
\sqrt{\frac{\log \lambda_{\max}\{N_m\}}{\lambda_{\min}\{N_m\}}}.\label{Eq43}
\end{align}

From (\ref{Eq37}), it follows that for some $c_4>0$,
\begin{align}
\nonumber&\left|\mu^{(2)T}_{N_m+1}\Phi^{(21)}_{N_m}\mu^{(1)}_{N_m+1}\right|\\
\nonumber\leq&\left\|\mu^{(2)}_{N_m+1}\right\|\left\|\Phi^{(21)}_{N_m}\right\|\left\|\mu^{(1)}_{N_m+1}\right\|
\leq \left\|\mu^{(2)}_{N_m+1}\right\|\cdot\left\|\mu^{(1)}_{N_m+1}\right\|\cdot c_4\left\|\Phi^{(21)}_{N_m}\right\|_F\\
\nonumber\leq&\left\|\mu^{(2)}_{N_m+1}\right\|\cdot\left\|\mu^{(1)}_{N_m+1}\right\|\cdot c_4\left\|\Phi_{N_m}\right\|_F\leq \left\|\mu^{(2)}_{N_m+1}\right\|\cdot\left\|\mu^{(1)}_{N_m+1}\right\|\cdot c_4\left\|\Phi_{N_m}\right\|\\
\leq& c_4\lambda_{\max}(N_m)\left\|\mu^{(2)}_{N_m+1}\right\| \left(\frac{\lambda_{\max}(N_m)}{\lambda_{\min}(N_m)}\sqrt{\frac{\log \lambda_{\max}(N_m)}{\lambda_{\min}(N_m)}}
+\frac{\lambda_{N_m}}{\lambda_{\min}(N_m)}\right).\label{Eq44}
\end{align}

From the definition of $\widehat{\theta}_{N_m+1}(l)$,
for some $c_6>c_5>0$,
\begin{align}
c_5\sqrt{\frac{\log \lambda_{\max}(N_m)}{\lambda_{\min}(N_m)}}  \leq\left|\widehat{\theta}_{N_m+1}(l)\right|\leq c_6\sqrt{\frac{\log \lambda_{\max}(N_m)}{\lambda_{\min}(N_m)}},~~l=d+1,\dots,r\label{Re45}
\end{align}
and
\begin{align}
\frac{1}{c_6\sqrt{\frac{\log \lambda_{\max}(N_m)}{\lambda_{\min}(N_m)}}  } \leq\frac{1}{\left|\widehat{\theta}_{N_m+1}(l)\right|}\leq \frac{1}{c_5\sqrt{\frac{\log \lambda_{\max}(N_m)}{\lambda_{\min}(N_m)}}},~~l=d+1,\dots,r\label{Re46}
\end{align}
and hence for some $c_7>0$
\begin{align}
\nonumber&\lambda_{N_m}\sum\limits_{l=d+1}^r\frac{1}{|\widehat{\theta}_{N_m+1}(l)|}|\mu_{N_m+1}(l)|\\
\nonumber\geq & c_7\lambda_{N_m}\frac{1}{\sqrt{\frac{\log \lambda_{\max}(N_m)}{\lambda_{\min}(N_m)}}} \sum\limits_{l=d+1}^r|\mu_{N_m+1}(l)|\\
\geq & c_7\frac{\lambda_{N_m}}{\sqrt{\frac{\log \lambda_{\max}(N_m)}{\lambda_{\min}(N_m)}}} \|\mu^{(2)}_{N_m+1}\|.\label{Eq47}
\end{align}

From (\ref{Eq41}), (\ref{Eq43}), (\ref{Eq44}), and (\ref{Eq47}), we obtain
\begin{align}
\nonumber&J_{N_m+1}(\theta+\mu_{N_m+1})-J_{N_m+1}(\theta+\overline{\mu}_{N_m+1})\\
\nonumber\geq & \lambda_{\min}(N_m)\|\mu^{(2)}_{N_m+1}\|^2-c_3\lambda_{\max}(N_m) \left\|\mu^{(2)}_{N_m+1}\right\|
\sqrt{\frac{\log \lambda_{\max}\{N_m\}}{\lambda_{\min}\{N_m\}}}\\
\nonumber&-c_4\lambda_{\max}(N_m)\left\|\mu^{(2)}_{N_m+1}\right\| \left(\frac{\lambda_{\max}(N_m)}{\lambda_{\min}(N_m)}\sqrt{\frac{\log \lambda_{\max}(N_m)}{\lambda_{\min}(N_m)}}
+\frac{\lambda_{N_m}}{\lambda_{\min}(N_m)}\right)\\
\nonumber&+c_7\frac{\lambda_{N_m}}{\sqrt{\frac{\log \lambda_{\max}(N_m)}{\lambda_{\min}(N_m)}}} \|\mu^{(2)}_{N_m+1}\|\\
\nonumber=&\lambda_{\min}(N_m)\|\mu^{(2)}_{N_m+1}\|\Bigg[\|\mu^{(2)}_{N_m+1}\| -c_3\frac{\lambda_{\max}\{N_m\}}{\lambda_{\min}\{N_m\}}\sqrt{\frac{\log \lambda_{\max}(N_m)}{\lambda_{\min}(N_m)}}\\
\nonumber&-c_4\frac{\lambda_{\max}(N_m)}{\lambda_{\min}(N_m)} \left(\frac{\lambda_{\max}(N_m)}{\lambda_{\min}(N_m)}\sqrt{\frac{\log \lambda_{\max}(N_m)}{\lambda_{\min}(N_m)}}
+\frac{\lambda_{N_m}}{\lambda_{\min}(N_m)}\right)\\
&+c_7\frac{\lambda_{N_m}}{\lambda_{\min}(N_m)\sqrt{\frac{\log \lambda_{\max}(N_m)}{\lambda_{\min}(N_m)}}}\Bigg].\label{Eq48}
\end{align}

By assumption A4), it follows that for any $\varepsilon>0$, there exists $M_0>0$ large enough such that for any $m\geq M_0$,
\begin{align}
\frac{\lambda_{\max}(N_m)}{\lambda_{\min}(N_m)}\sqrt{\frac{\log \lambda_{\max}(N_m)}{\lambda_{\min}(N_m)}}
\leq \varepsilon\frac{\lambda_{N_m}}{\lambda_{\min}(N_m)}\leq \varepsilon\frac{\lambda_{N_m}}{\lambda_{\min}(N_m)\sqrt{\frac{\log \lambda_{\max}(N_m)}{\lambda_{\min}(N_m)}}}\label{Eq49}
\end{align}
and hence
\begin{align}
\nonumber&\frac{\lambda_{\max}(N_m)}{\lambda_{\min}(N_m)} \left(\frac{\lambda_{\max}(N_m)}{\lambda_{\min}(N_m)}\sqrt{\frac{\log \lambda_{\max}(N_m)}{\lambda_{\min}(N_m)}}+\frac{\lambda_{N_m}}{\lambda_{\min}(N_m)}\right)\\
\leq&(1+\varepsilon)\frac{\lambda_{\max}(N_m)}{\lambda_{\min}(N_m)}\frac{\lambda_{N_m}}{\lambda_{\min}(N_m)}. \label{Eq50}
\end{align}

By assumption A3), we have
\begin{align}
\frac{\frac{\lambda_{\max}(N_m)}{\lambda_{\min}(N_m)}\frac{\lambda_{N_m}}{\lambda_{\min}(N_m)}} {\frac{\lambda_{N_m}}{\lambda_{\min}(N_m)\sqrt{\frac{\log \lambda_{\max}(N_m)}{\lambda_{\min}(N_m)}}}}
=\frac{\lambda_{\max}(N_m)}{\lambda_{\min}(N_m)}\sqrt{\frac{\log \lambda_{\max}(N_m)}{\lambda_{\min}(N_m)}}
=o(1)~~\mathrm{as}~m\to\infty.\label{Eq51}
\end{align}

From (\ref{Eq50}) and (\ref{Eq51}),
 we have that for all $m$ large enough
\begin{align}
&\frac{\lambda_{\max}(N_m)}{\lambda_{\min}(N_m)} \left(\frac{\lambda_{\max}(N_m)}{\lambda_{\min}(N_m)}\sqrt{\frac{\log \lambda_{\max}(N_m)}{\lambda_{\min}(N_m)}}+\frac{\lambda_{N_m}}{\lambda_{\min}(N_m)}\right) \leq\varepsilon(1+\varepsilon) \frac{\lambda_{N_m}}{\lambda_{\min}(N_m)\sqrt{\frac{\log \lambda_{\max}(N_m)}{\lambda_{\min}(N_m)}}}. \label{Eq52}
\end{align}

By (\ref{Eq48}), (\ref{Eq49}), and (\ref{Eq52}), we obtain
\begin{align}
\nonumber0\geq &J_{N_m+1}(\theta+\mu_{N_m+1})-J_{N_m+1}(\theta+\overline{\mu}_{N_m+1})\\
\nonumber\geq&\lambda_{\min}(N_m)\|\mu^{(2)}_{N_m+1}\|\Bigg[\|\mu^{(2)}_{N_m+1}\| +\big(-\varepsilon-\varepsilon(1+\varepsilon)\big)c_8\frac{\lambda_{N_m}}{\lambda_{\min}(N_m)\sqrt{\frac{\log \lambda_{\max}(N_m)}{\lambda_{\min}(N_m)}}}\\
&+c_7\frac{\lambda_{N_m}}{\lambda_{\min}(N_m)\sqrt{\frac{\log \lambda_{\max}(N_m)}{\lambda_{\min}(N_m)}}}\Bigg]\label{Eq53}
\end{align}
where $c_8>0$ is a constant.

Note
that $\mu_{N_m+1}(i_l)\neq0,~i_l\in\{d+1,\dots,r\}$ and hence $\|\mu^{(2)}_{N_m+1}\|>0$. Since $\varepsilon$ in (\ref{Eq53}) can be sufficiently small such that $-(\varepsilon+\varepsilon(1+\varepsilon))c_8+c_7>0$,
  $J_{N_m+1}(\theta+\mu_{N_m+1})-J_{N_m+1}(\theta+\overline{\mu}_{N_m+1})>0$. The contradiction with (\ref{Eq38}) indicates that $\|\mu^{(2)}_{N+1}\|=0$ for all $N$ large enough and hence (\ref{Eq16}) holds. This finishes the proof.
\hfill$\square$

\subsection{Comparison of Conditions for Algorithms (\ref{Eq5})--(\ref{Eq8}) to That of Persistent Excitation, Information Criteria for Order Estimation and LASSO}

From Theorem \ref{Thm1}, we can find that for consistency of sparse identification algorithm for system (\ref{Eq1}), an essential requirement on the observation data is assumption A3), which includes the classical persistent excitation (PE) condition (e.g., \cite{Ljung87}) as its special case.
 That is,
  if $\frac{\lambda_{\max}(N)}{\lambda_{\min}(N)}=O(1)$, then
\begin{align}
\frac{\lambda_{\max}(N)}{\lambda_{\min}(N)}\sqrt{\frac{\log \lambda_{\max}(N)}{\lambda_{\min}(N)}}=O\left(\sqrt{\frac{\log \lambda_{\max}(N)}{\lambda_{\min}(N)}}\right)
    \mathop{\longrightarrow}\limits_{N\to\infty}0~~\mathrm{a.s.,}\label{Eq54}
\end{align}
and in this case, the coefficient $\{\lambda_N\}$ in algorithms (\ref{Eq5})--(\ref{Eq8}) can be chosen as $\lambda_N=\lambda_{\min}(N)^{\frac{1}{2}+\epsilon}$ for any fixed $\epsilon\in(0,\frac{1}{2})$, which meets the requirements in assumption A4), i.e.,
\begin{align}
&\frac{\lambda_N}{\lambda_{\min}(N)}= O\left(\frac{\lambda_{\min}(N)^{\frac{1}{2}+\epsilon}}{\lambda_{\min}(N)}\right) \mathop{\longrightarrow}\limits_{N\to\infty}0,\label{Eq55}\\
&\frac{\lambda_{\max}(N)\sqrt{\frac{\log \lambda_{\max}(N)}{\lambda_{\min}(N)}}}{\lambda_N}=O\left(\frac{\lambda_{\min}(N)\sqrt{\frac{\log \lambda_{\min}(N)}{\lambda_{\min}(N)}}}{\lambda_{\min}(N)^{\frac{1}{2}+\epsilon}}\right) \mathop{\longrightarrow}\limits_{N\to\infty}0.\label{Eq56}
\end{align}

Compared with the
celebrated order estimation methods for stochastic systems,  for example,
Akaike's information criterion (AIC)\cite{Akaike}, Bayesian information criterion (BIC) \cite{HannanDiestler}, control information criteria (CIC) \cite{ChenGuo}, etc., the sparse identification algorithms given in this paper, in fact, go further, i.e., once the sets of zero and nonzero elements in the parameter vector being correctly identified, the estimates for system order follow directly;  see Table \ref{tab1} for a detailed comparison.
\begin{table*}[htbp!]
\caption{Comparison between Order Estimation Methods and Algorithms (\ref{Eq5})--(\ref{Eq8}) ($\surd$ indicates yes and $\bigcirc$ for no)}
\begin{center}
\begin{tabular}{|c c c c c|}
\hline & Stationary Time Series & Closed-loop Systems & $\begin{array}{c}\mbox{Estimation~for}\\ \mbox{System~Order}\end{array}$ & $\begin{array}{c}\mbox{Estimation~for}\\ \mbox{Sparse~Parameters}\end{array}$\\
\hline AIC, BIC & $\surd$ & $\bigcirc$ & $\surd$ & $\bigcirc$ \\
\hline CIC & $\surd$ & $\surd$ & $\surd$ & $\bigcirc$ \\
\hline Algorithms (\ref{Eq5})--(\ref{Eq8})  & $\surd$ & $\surd$ & $\surd$ & $\surd$\\
\hline
\end{tabular}
\end{center}
\label{tab1}
\end{table*}

Next, we make a comparison between assumption A3) and the classical regular and irrepresentable conditions for consistency of LASSO and its variants (\cite{ZhaoYu} \cite{Zou}). For simplicity of notations, we still assume that the parameter vector $\theta=[\theta_1^T~\theta_2^T]^T$, $\theta_1=[\theta(1)\dots \theta(d)]^T$, $\theta_2=[\theta(d+1)\dots \theta(r)]^T$ such that $\theta(i)\neq 0,~i=1,\dots,d$ and $\theta(j)=0,~j=d+1,\dots,r$.

Denote
\begin{align*}
C_N\triangleq\frac{1}{N}\sum\limits_{k=1}^N\varphi_k\varphi_k^T=
\begin{bmatrix}C_N^{11} & C_N^{12}\\C_N^{21} & C_N^{22}\end{bmatrix}
\end{align*}
where $C_N^{11} \in \mathbb{R}^{d\times d}$ and $C_N^{12},~C_N^{21}$ and $C_N^{22}$ are with compatible dimensions. The comparison on conditions for consistency of LASSO and its variations and algorithms (\ref{Eq5})--(\ref{Eq8}) is made in Table \ref{tab2}.

\begin{table*}[htbp!]
\caption{Conditions for Consistency of LASSO and Its Variations and Algorithms (\ref{Eq5})--(\ref{Eq8})}
\begin{center}
\begin{tabular}{|c|c|}
\hline & Conditions on System\\
\hline LASSO (\cite{ZhaoYu}) & \begin{tabular}{c}Regularity~Condition:~$C_N\to C>0~\mbox{as}~N\to\infty$\\ Strong~Irrepresentable~Condition:
$\mbox{ for some }\eta>0,~ |C_N^{21}(C_N^{11})^{-1}\mathrm{sgn}(\theta_1)|\leq1-\eta$\end{tabular}\\
\hline Adaptive LASSO (\cite{Zou}) & $\begin{array}{l}\mbox{Regularity Condition:}~C_N\to C>0~\mbox{as}~N\to\infty\end{array}$\\
\hline Algorithms (\ref{Eq5})--(\ref{Eq8}) & $
    \frac{\lambda_{\max}(N)}{\lambda_{\min}(N)}\sqrt{\frac{\log \lambda_{\max}(N)}{\lambda_{\min}(N)}}
    \mathop{\longrightarrow}\limits_{N\to\infty}0
    $\\
\hline
\end{tabular}
\end{center}
\label{tab2}
\end{table*}

Here in the {\em strong irrepresentable condition} given in Table \ref{tab2} the $\mathrm{sgn}(\cdot)$ function as well as the inequality are understood in the element-wise sense. From Table \ref{tab2} it is to directly verify that assumption A3) given in this paper includes the {\em regularity condition} as its special case and the strong irrepresentable condition, which adopts {\em a prior} structural information on sparsity of the parameter vector, is not required.

\section{Application to Identification of Hammerstein Systems and Linear Stochastic Systems with Self-tuning Regulation Control}

\subsection{Application to Basis Function Selection of Hammerstein Systems}

The Hammerstein system is a block-oriented nonlinear system consisting of a static nonlinear function followed by a linear dynamic. This kind of nonlinear systems is widely applied in modelling the complicated realistic phenomena such as distillation columns \cite{Eskinat}, power amplifier \cite{Kim}, etc.

We consider a Hammerstein system with its linear subsystem being an ARX system and the nonlinear function being a combination of basis functions with unknown coefficients:
\begin{align}
&y_{k+1}=a_1y_k+\cdots+a_py_{k+1-p}+b_1f(u_k)+\cdots+b_qf(u_{k+1-q})+w_{k+1},\label{Eq57}\\
&f(u_k)=\sum\limits_{j=1}^sd_jg_j(u_k),\label{Eq58}
\end{align}
where $\{g_j(\cdot)\}_{j=1}^s$ are the basis functions.
The identification task of system (\ref{Eq57}) is to estimate the parameters $\{a_i,b_j\}$ of the linear subsystem and the coefficients $\{d_l\}$ in the nonlinear function. In practice, the system representation (\ref{Eq57})--(\ref{Eq58}) is likely to be sparse. First the system is unknown and the assumed order of the linear part has to be high. Further, to model the unknown nonlinear part, the number of nonlinear terms has to be large.

By setting
\begin{align}
&\theta_H=[a_1\dots a_p~(b_1d_1)\dots (b_1d_s)\dots (b_qd_1)\dots (b_qd_s)]^T,\label{Eq59}\\
&\varphi_{k,H}=[y_k\dots y_{k+1-p}~g_1(u_k)\dots g_s(u_k)\dots g_1(u_{k+1-q})\dots g_s(u_{k+1-q})]^T,\label{Eq60}
\end{align}
the Hammerstein system is written in a compact form
\begin{align}
y_{k+1}=\theta_H^T\varphi_{k,H}+w_{k+1}.\label{Eq61}
\end{align}
Thus the estimates for $\{a_i,b_j,d_l\}$ can be derived by identifying the vector $\theta_H$. This is called the over-parametrization method in literature \cite{Bai}\cite{Chaoui}.

For
Hammerstein system (\ref{Eq57}), in order to well approximate the nonlinear function $f(\cdot)$ it usually adopts a large number of basis functions, which sometimes leads to a redundant representation of the system and the high dimensionality of $\theta_H$.
To obtain a simple but precise model of the system, it is natural to ask how to determine the effective basis functions in $\{g_j(\cdot)\}_{j=1}^s$, or equivalently, the sparse identification of the parameter vector $\theta_H$.
Note
that the linear regression form (\ref{Eq61}) coincides with (\ref{Eq1}). Thus algorithms (\ref{Eq5})--(\ref{Eq8}) can be applied.

Denote
\begin{align}
M\triangleq
\begin{bmatrix}
b_1d_1 & \cdots & b_1d_s\\
\vdots & \ddots & \vdots\\
b_qd_1 & \cdots & b_qd_s
\end{bmatrix}
=
\begin{bmatrix}
M(1) & \cdots & M(s)
\end{bmatrix}\label{Eq62}
\end{align}
with $M(l)=[b_1d_l\cdots b_qd_l]^T,~l=1,\dots,s$. So the noneffective basis functions in $\{g_j(\cdot)\}_{j=1}^s$ correspond to zero columns in matrix $M$.

Denote the LS estimate for the Hammerstein system by $\theta_{N+1,H}$. Before presenting the results, we need the following assumptions.

\begin{itemize}
\item[B1)] $A(z)=1-a_1z-\cdots-a_pz^p$ is stable, i.e., $|A(z)|\neq 0,~\forall|z|\leq 1$ and $b_1^2+\cdots+b_q^2\neq 0$.
\item[B2)] $\{1,g_1(x),\dots,g_s(x)\}$ is linearly independent over some interval $[a,b]$.
\item[B3)] $\{u_k\}_{k\geq1}$ is an i.i.d. sequence with density $p(x)$ which is positive and continuous on $[a,b]$ and $0<\mathbb{E}g^2_i(u_k)<\infty,~i=1,\dots,s$. Further, $\{u_k\}_{k\geq1}$ and $\{w_k\}_{k\geq1}$ are mutually independent.
\end{itemize}

\begin{proposition}\label{Prop1}(\cite{Zhao2010})
If A1) and B1)-B3) hold, then for the maximal and minimal eigenvalues of $\sum_{k=1}^N\varphi_{k,N}\varphi^T_{k,N}$, it holds that
\begin{align}
&c_1N\leq\lambda_{\max}\left\{\sum\limits_{k=1}^N\varphi_{k,H}\varphi^T_{k,H}\right\}\leq c_2N,~~\mathrm{a.s.}\label{Eq63}\\
&c_3N\leq\lambda_{\min}\left\{\sum\limits_{k=1}^N\varphi_{k,H}\varphi^T_{k,H}\right\}\leq c_4N,~~\mathrm{a.s.,}\label{Eq64}
\end{align}
for some $0<c_1<c_2$, $0<c_3<c_4$, and for the LS estimate $\theta_{N+1,H}$,
\begin{align}
\|\theta_{N+1,H}-\theta_H\|=O\left(\sqrt{\frac{\log N}{N}}\right)~~\mathrm{a.s.}\label{Eq65}
\end{align}
\end{proposition}

Then from $\{\theta_{N+1,H}\}$ and by algorithms (\ref{Eq5})--(\ref{Eq8}), we can have the sparse estimates $\{\beta_{N+1,H}\}$ for the parameters in the Hammersten system. Denote
\begin{align}
&\beta_{N+1,H}=[a_{1,N+1}\dots a_{p,N+1}~(b_1d_1)_{N+1}\dots (b_1d_s)_{N+1}\dots (b_qd_1)_{N+1}\dots (b_qd_s)_{N+1}]^T,\label{Eq66}\\
&M_{N+1}=[M_{N+1}(1)\cdots M_{N+1}(s)],\label{Eq67}
\end{align}
with $M_{N+1}(l)=[(b_1d_l)_{N+1}\cdots (b_qd_l)_{N+1}]^T,~l=1,\dots,s$ and
\begin{align}
&B^*=\{l=1,\cdots,s~\Big|~d_l=0\},\label{Eq68}\\
&B^*_{N+1}=\{l=1,\dots,s~\Big|~M_{N+1}(l)=0\}.\label{Eq69}
\end{align}

\begin{proposition}\label{Prop2}
Set $\lambda_N=N^{\frac{1}{2}+\epsilon}$ for fixed $\epsilon\in(0,\frac{1}{2})$. If A1) and B1)-B3) hold for Hammerstein system (\ref{Eq57})--(\ref{Eq58}), then there exists an $\omega$-set $\Omega_0$ with $\mathbb{P}\{\Omega_0\}=1$ such that for any $\omega\in\Omega_0$, there exists an integer $N_0(\omega)$ such that
\begin{align}
B^*_{N+1}=B^*,~~\forall~N\geq N_0(\omega),\label{Eq70}
\end{align}
i.e., the effective basis functions in $\{g_j(\cdot)\}_{j=1}^s$ can be correctly identified.
\end{proposition}

{\em Proof: }
By (\ref{Eq63}), (\ref{Eq64}) and noticing $\lambda_N=N^{\frac{1}{2}+\epsilon}$, $\epsilon\in(0,\frac{1}{2})$, we can verify that A1)--A4) hold for the regression model (\ref{Eq61}) and by Theorem \ref{Thm1}, the results follow directly.
\hfill$\square$

\begin{remark}\label{Remark5}
By
noting that
$$
M\triangleq
\begin{bmatrix}
b_1\\
\vdots\\
b_q
\end{bmatrix}
\begin{bmatrix}
d_1 & \cdots & d_s
\end{bmatrix},
$$
we can further obtain the estimates for the nonzero elements in $\{b_i,~i=1,\cdots,q\}$ and $\{d_l,~l=1,\cdots,s\}$ by performing a singular value decomposition (SVD) algorithm to $M_{N+1}$ defined by (\ref{Eq67}); see \cite{Bai} and \cite{Chaoui} for details.
\end{remark}

\subsection{Application to Sparse Parameter Estimation of Linear Stochastic Systems with Self-tuning Regulation Control}

In the above section, the observation data are collected from an open-loop Hammerstein system. In this section, we apply algorithms (\ref{Eq5})--(\ref{Eq8}) to the sparse parameter estimation of a closed-loop system. The self-tuning regulation (STR) control, first proposed by {\AA}str\"{o}m and Wittenmark \cite{Astrom1973} in 1973, has received much attention from theoretical research and has been successfully applied in industrial practice. Briefly speaking, the goal of STR is to minimize the tracking error of the system with unknown parameters, which clearly consists of a closed-loop system.


Let us consider
a one-dimensional ARX system:
\begin{align}
y_{k+1}=a_1y_k+\cdots+a_py_{k+1-p}+b_1u_k+\cdots+b_qu_{k+1-1}+w_{k+1},\label{Eq71}
\end{align}
where, using
the same notations as in previous sections, $u_k$, $y_k$, and $w_k$ are the system input, output, and noise, respectively, and $\{a_i,b_j\}$ are the unknown parameters.
Denote
\begin{align*}
&A(z)=1-a_1z-\cdots-a_pz^p,\\
&B(z)=b_1+b_2z+\cdots+b_qz^{q-1},\\
&\theta_L=[a_1,\dots, a_p,b_1,\dots, b_q]^T,\\
&\varphi_{k,L}=[y_k,\dots, y_{k+1-p},u_k,\dots, u_{k+1-q}]^T.
\end{align*}
Then system (\ref{Eq71}) can directly be formulated into a linear regression form as system (\ref{Eq1}). Let $\{y_k^*\}$ be a sequence of deterministic bounded reference signals. The problem is to guarantee the optimal tracking performance of the closed-loop system and meanwhile, to correctly identify the sets of zero and nonzero elements in $\theta_L$.


Denote the LS estimate for vector $\theta_L$ in the ARX system by $\theta_{k,L}$. Since $\theta_L$ is unknown, the {\em Certainty Equivalence Principle} (\cite{Astrom1973}\cite{Guo1991}) suggests to define the adaptive control $u^0_k$ from
\begin{align}
\theta_{k,L}^T\varphi_k=y_{k+1}^*\label{Eq72}
\end{align}
or equivalently,
\begin{align}
u^0_k=\frac{1}{b_{1,k}}\{y_{k+1}^*+(b_{1,k}u_k-\theta_{k,L}^T\varphi_k)\}\label{Eq73}
\end{align}
where $\theta_{k,L}$ and $b_{1,k}$ are the LS estimates for $\theta_L$ and $b_1$, respectively.

For identification of the closed-loop system, some excitation on the system is required. In order that the external excitation does not worsen the control performance of STR, the diminishing excitation technique is applied \cite{Guo1991}.
Let $\{w'_k\}$ be an i.i.d. and bounded random sequence with $\mathbb{E}w'_k=0,~\mathbb{E}(w'_k)^2=1$. Based on the control input $u^0_k$ defined by (\ref{Eq73}), the diminishing excitation input $u_k$ is defined as
\begin{align}
u_k=u^0_k+\frac{w'_k}{r_{k-1}^{\overline{\varepsilon}/2}},~~k\geq1\label{Eq76}
\end{align}
with $r_{k-1}=1+\sum\limits_{i=1}^{k-1}\|\varphi_{i,L}\|^2$ and $\overline{\varepsilon}\in(0,\frac{1}{2(t+1)}),~t=\max\{p,q\}+p-1$. Then $u_k$ serves as the system input at time $k$.

The assumptions made on (\ref{Eq71}) are as follow:
\begin{itemize}
\item[C1)] The noise
$\{w_k,\mathcal{F}_k\}_{k\geq1}$ is a martingale difference sequence, i.e., $\mathbb{E}[w_{k+1}|\mathcal{F}_k]=0,~k\geq1$, and there exists some $\gamma>2$ such that $\sup\limits_k\mathbb{E}\big[|w_{k+1}|^{\gamma}|\mathcal{F}_k\big]<\infty~\mathrm{a.s.}$ and
    $$
    \lim\limits_{k\to\infty}\frac{1}{k}\sum\limits_{i=1}^kw^2_i=R>0~~\mathrm{a.s.}
    $$
\item[C2)] $B(z)\neq 0,~\forall~z:~|z|\leq1$.
\item[C3)] $|a_p|+|b_q|\neq0$.
\end{itemize}

Assumption C2) is usually called the minimum phase condition. Since $b_{1,k}$ is in the denominator of (\ref{Eq73}), we further impose the following assumption.
\begin{itemize}
\item[C4)] $u_k$ is well-defined from (\ref{Eq72}) or (\ref{Eq73}) for each $k\geq0$.
\end{itemize}



%

The following result, i.e., the stability and optimality of STR, is well known
in literature.

\begin{proposition}\label{Prop4}(\cite{Guo1991})
Assume that C1)--C4) hold. Then the STR with diminishing excitation is stable and optimal, i.e.,
\begin{align}
&\limsup\limits_{k\to\infty}\frac{1}{k}\sum\limits_{i=1}^k(\|u_i\|^2+\|y_i\|^2)<\infty~~\mathrm{a.s.}\label{Eq74}\\
&\lim\limits_{k\to\infty}\frac{1}{k}\sum\limits_{i=1}^k(y_i-y_i^*)^2=R~~\mathrm{a.s.}\label{Eq75}
\end{align}
and the LS estimates $\{\theta_{N+1,L}\}$ are strongly consistent, and further,
\begin{align}
&\lambda_{\max}\left\{\sum\limits_{k=1}^N\varphi_{k,L}\varphi_{k,L}^T\right\}=O(N),\label{Eq77}\\
&\lambda_{\min}\left\{\sum\limits_{k=1}^N\varphi_{k,L}\varphi_{k,L}^T\right\}\geq cN^{1-\overline{\varepsilon}(t+1)},\label{Eq78}
\end{align}
for some $c>0$ which may depend on sample paths and $\overline{\varepsilon}>0$ specified in (\ref{Eq76}).
\end{proposition}

From Proposition \ref{Prop4}, we find that the regularity condition for consistency of LASSO (see, e.g., Table \ref{tab2}) may not take place for the closed-loop system.
Then from $\{\theta_{N+1,L}\}$ and by algorithms (\ref{Eq5})--(\ref{Eq8}), we can have the sparse estimates $\{\beta_{N+1,L}\}$ for the parameters in the ARX system. Denote $\beta_{N,L}\triangleq[\beta_{N,L}(1)\dots \beta_{N,L}(p+q)]^T$ and
\begin{align}
&C^*\triangleq\{i=1,\cdots,p;~j=1,\dots,q~|~a_i=0,~b_j=0\}\label{Eq79}\\
&C^*_N\triangleq\{i=1,\dots,p+q~|~\beta_{N,L}(i)=0\}.\label{Eq80}
\end{align}

Based on Proposition \ref{Prop4}, for the estimate $\beta_{N+1,L}$ generated from (\ref{Eq5})--(\ref{Eq8}) with data from closed-loop system (\ref{Eq71})--(\ref{Eq76}), we have the following result.

\begin{proposition}\label{Prop5}
Set the parameter $\overline{\varepsilon}$ in the diminishing excitation satisfying $\overline{\varepsilon}(t+1)\in(0,\frac{1}{4})$ and the coefficient in algorithm (\ref{Eq8}) as $\lambda_N=N^{1-\frac{3}{2}\overline{\varepsilon}(1+t)}$. If C1)--C4) hold for closed-loop system (\ref{Eq71})--(\ref{Eq76}), then there exists an $\omega$-set $\Omega_0$ with $\mathbb{P}\{\Omega_0\}=1$ such that for any $\omega\in\Omega_0$, there exists an integer $N_0(\omega)$ such that
\begin{align}
C^*_{N+1}=C^*,~~\forall~N\geq N_0(\omega),\label{Eq81}
\end{align}
i.e., the zero and nonzero elements in $\theta_L$ can be correctly identified.
\end{proposition}

{\em Proof: }
By Theorem \ref{Thm1}, we only need to verify that assumptions A3) and A4) hold true for the closed-loop system and the specified coefficient $\lambda_N$.

By Proposition \ref{Prop4}, it immediately follows that
\begin{align}
\nonumber&\frac{\lambda_{\max}(N)}{\lambda_{\min}(N)}\sqrt{\frac{\log \lambda_{\max}(N)}{\lambda_{\min}(N)}}=
O\left(\frac{N}{N^{1-\overline{\varepsilon}(t+1)}}\sqrt{\frac{\log N}{N^{1-\overline{\varepsilon}(t+1)}}}\right)\\
=&O\left(\frac{N^{\overline{\varepsilon}(t+1)}}{N^{1/2-\overline{\varepsilon}(t+1)/2}}\sqrt{\log N}\right)
=O\left(\frac{1}{N^{\frac{1}{2}-\frac{3}{2}\overline{\varepsilon}(t+1)}}\sqrt{\log N}\right)
\mathop{\longrightarrow}\limits_{N\to\infty}0\label{Eq82}
\end{align}
by noting
that $\overline{\varepsilon}(t+1)\in(0,\frac{1}{4})$. Hence assumption A3) holds.

By noting
$\lambda_N=N^{1-\frac{3}{2}\overline{\varepsilon}(t+1)}$, we have
\begin{align}
\frac{\lambda_N}{\lambda_{\min}(N)} =O\left(\frac{N^{1-\frac{3}{2}\overline{\varepsilon}(t +1)}}{N^{1-\overline{\varepsilon}(t+1)}}\right) \mathop{\longrightarrow}\limits_{N\to\infty}0\label{Eq83}
\end{align}
and
\begin{align}
\nonumber&\frac{\lambda_{\max}(N)\sqrt{\frac{\log \lambda_{\max}(N)}{\lambda_{\min}(N)}}}{\lambda_N}
=
O\left(\frac{N}{N^{1-\frac{3}{2}\overline{\varepsilon}(t+1)}}\sqrt{\frac{\log N}{N^{1-\overline{\varepsilon}(t+1)}}}\right)\\
=&O\left(\frac{1}{N^{\frac{1}{2}-2\overline{\varepsilon}(t+1)}}\sqrt{\log N}\right)
\mathop{\longrightarrow}\limits_{N\to\infty}0\label{Eq84}
\end{align}
by noting
that $\overline{\varepsilon}(t+1)\in(0,\frac{1}{4})$. Hence assumption A4) holds.
By applying Theorem \ref{Thm1},
(\ref{Eq81}) holds true.
\hfill$\square$

\section{Simulation}

In this section, we consider two examples, one being an open-loop system and the other being a closed-loop system with self-tuning regulation control, to testify the performance of the identification algorithms (\ref{Eq5})--(\ref{Eq8}).

{\em Example I.} Consider the following Hammerstein system,
\begin{align*}
y_{k+1}+a_1y_k+a_2y_{k-1}=b_1f(u_k)+b_2f(u_{k-1})+w_{k+1},
\end{align*}
where $a_1=-1.5,~a_2=0.56,~b_1=1,~b_2=-2$ and $f(u)=\sum_{j=1}^6d_ju^j$ is a $6$-th polynomial with $d_1=1,~d_3=0.2,~d_5=0.009$ and $d_2=d_4=d_6=0$. Denote
\begin{align*}
&\theta=[-a_1~-a_2~(b_1d_1)\cdots(b_1d_6)~(b_2d_1)\cdots(b_2d_6)]^T\\
&\varphi_k=[y_k~y_{k-1}~u_k\cdots u_{k}^6~u_{k-1}\cdots u_{k-1}^6]^T.
\end{align*}
It is to directly verify that the Hammerstein system can be formulated by $y_{k+1}=\theta^T\varphi_k+w_{k+1}$ and the following equality takes place
\begin{align*}
\begin{bmatrix}
a_1 & b_1d_1 & b_1d_2 & b_1d_3 & b_1d_4 & b_1d_5 & b_1d_6 \\
a_2 & b_2d_1 & b_2d_2 & b_2d_3 & b_2d_4 & b_2d_5 & b_2d_6
\end{bmatrix}
=
\begin{bmatrix}
-1.5 & 1 & 0 & 0.2 & 0 & 0.009 & 0 \\
0.56 & -2 & 0 & -0.4 & 0 & -0.018 & 0
\end{bmatrix}.
\end{align*}

For identification of the Hammerstein system, we select the input $\{u_k\}$ as an i.i.d. sequence that is uniformly distributed over $[-5,5]$. We assume that the noise sequence $\{w_k\}$ is iid with Gaussian distribution $\mathcal{N}(0,1)$ and independent of $\{u_k\}$.

\begin{figure}[tbp]
\centering
\includegraphics[width=16cm,trim=100 0 100 0, clip]{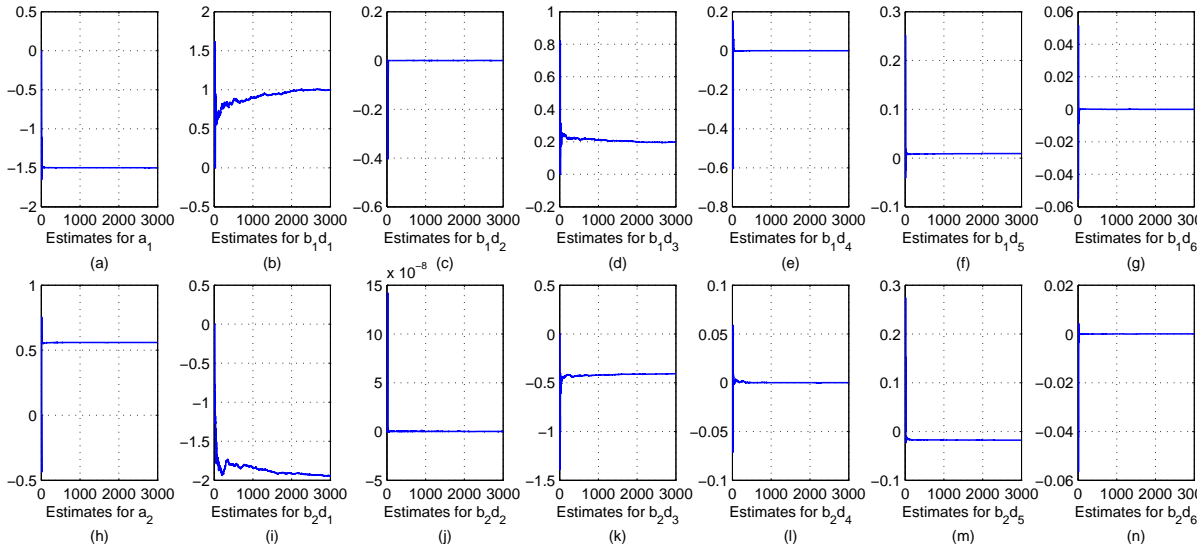}
\caption{Estimation Sequences $\{a_{1,N},a_{2,N},(b_1d_1)_N,\cdots,(b_1d_6)_N,(b_2d_1)_N,\cdots,(b_2d_6)_N\}_{N=1}^{3000}$}\label{Fig2}
\end{figure}

Figure \ref{Fig2} shows the estimation sequences $\{a_{1,N},a_{2,N},(b_1d_1)_N,\cdots,(b_1d_6)_N,(b_2d_1)_N,\cdots,$ $(b_2d_6)_N\}_{N=1}^{3000}$ generated from algorithms (\ref{Eq5})--(\ref{Eq8}) with $\lambda_N=N^{0.75}$. Tables \ref{tab3} and \ref{tab4} compare the least squares estimates and the estimates generated from (\ref{Eq5})--(\ref{Eq8}) for $b_1d_2$, $b_1d_4$, $b_1d_6$, $b_2d_2$, $b_2d_4$, and $b_2d_6$, with different data length $N$. We adopt the Matlab CVX tools (\underline{http://cvxr.com/cvx/}) to solve the convex criterion (\ref{Eq8}). Although the optimization calculation procedure inevitably introduces numerical error, from Figure \ref{Fig2} and Tables \ref{tab3} and \ref{tab4}, we can find that, compared with the least squares estimates, algorithms (\ref{Eq5})--(\ref{Eq8}) generate sparser and more accurate estimates for the system parameters and thus give us valuable information in inferring the zero and nonzero elements in the unknown parameters. The simulation results are consistent with the theoretical analysis.

\begin{table*}[htbp!]\scriptsize
\caption{Comparison between Least Squares Estimates and Estimates Generated from (\ref{Eq5})--(\ref{Eq8})}
\begin{center}
\begin{tabular}{|c|c|c|c|c|c|}
\hline & N=100 & N=200 & N=300 & N=400 & N=500 \\

\hline \begin{tabular}{c}estimates for $b_1d_2$\\ from (\ref{Eq5})--(\ref{Eq8})\end{tabular} & $-6.3225\times 10^{-11}$ & $-4.9362\times 10^{-11}$ & $-2.9436\times 10^{-11}$ & $-1.4011\times 10^{-11}$ & $-9.2955\times 10^{-12}$ \\ \begin{tabular}{c}estimates for $b_1d_2$\\ by least squares\end{tabular} & $-0.0453$ & $-0.0076$ & $0.0234$ & $0.0236$ & $9.8015\times 10^{-4}$\\

\hline \begin{tabular}{c}estimates for $b_1d_4$\\ from (\ref{Eq5})--(\ref{Eq8})\end{tabular} & $-7.1882\times 10^{-4}$ & $-0.0015$ & $0.0086$ & $-6.5762\times 10^{-10}$ & $-1.8827\times 10^{-10}$ \\ \begin{tabular}{c}estimates for $b_1d_4$\\ by least squares\end{tabular} & $0.0082$ & $0.0049$ & $9.1888\times 10^{-4}$ & $7.9911\times 10^{-4}$ & $0.0028$\\

\hline \begin{tabular}{c}estimates for $b_1d_6$\\ from (\ref{Eq5})--(\ref{Eq8})\end{tabular} & $3.8980\times 10^{-5}$ & $7.8232\times 10^{-5}$ & $7.5117\times 10^{-5}$ & $8.3421\times\times 10^{-6}$ & $1.0843\times 10^{-5}$ \\ \begin{tabular}{c}estimates for $b_1d_6$\\ by least squares\end{tabular} & $-3.8050\times10^{-4}$ & $-2.6245\times10^{-4}$ & $-1.4367\times 10^{-4}$ & $-1.1328\times 10^{-4}$ & $-1.5350\times10^{-4}$\\
\hline
\end{tabular}
\end{center}
\label{tab3}
\end{table*}

\begin{table*}[htbp!]\scriptsize
\caption{Comparison between Least Squares Estimates and Estimates Generated from (\ref{Eq5})--(\ref{Eq8})}
\begin{center}
\begin{tabular}{|c|c|c|c|c|c|}
\hline & N=100 & N=200 & N=300 & N=400 & N=500 \\

\hline \begin{tabular}{c}estimates for $b_2d_2$\\ from (\ref{Eq5})--(\ref{Eq8})\end{tabular} & $-1.6121\times 10^{-11}$ & $-2.7542\times 10^{-11}$ & $2.3233\times 10^{-11}$ & $2.1635\times 10^{-11}$ & $3.7715\times 10^{-11}$ \\
\begin{tabular}{c}estimates for $b_2d_2$\\ by least squares\end{tabular} & $-0.0572$ & $-0.0138$ & $-0.0284$ & $-0.0268$ & $-0.0381$\\

\hline \begin{tabular}{c}estimates for $b_2d_4$\\ from (\ref{Eq5})--(\ref{Eq8})\end{tabular} & $0.0018$ & $5.8322\times10^{-4}$ & $0.0012$ & $3.1832\times 10^{-5}$ & $7.9625\times 10^{-10}$ \\
\begin{tabular}{c}estimates for $b_2d_4$\\ by least squares\end{tabular} & $0.0071$ & $0.0029$ & $0.0049$ & $0.0034$ & $0.0047$\\

\hline \begin{tabular}{c}estimates for $b_2d_6$\\ from (\ref{Eq5})--(\ref{Eq8})\end{tabular} & $-9.0533\times 10^{-5}$ & $-2.5922\times 10^{-5}$ & $-5.6134\times 10^{-5}$ & $-1.0984\times\times 10^{-6}$ & $-4.4105\times 10^{-6}$ \\
\begin{tabular}{c}estimates for $b_2d_6$\\ by least squares\end{tabular} & $-2.1416\times10^{-4}$ & $-1.0358\times10^{-4}$ & $-1.6503\times 10^{-4}$ & $-1.0778\times 10^{-4}$ & $-1.4685\times10^{-4}$\\
\hline
\end{tabular}
\end{center}
\label{tab4}
\end{table*}

{\em Example II.} Section IV.B establishes the consistent estimates for linear stochastic systems with sparse parameters under the self-tuning regulation control. Generally speaking, for a system which can be formulated into a linear regression form, the self-tuning regulation control can be applied. Let us consider the following Hammerstein system,
\begin{align}
y_{k+1}+a_1y_k=b_1(d_1u_k+d_2u_k^2+d_3u^3_k)+w_{k+1},\label{Eq85}
\end{align}
where $a_1=-0.5$, $b_1=2$, $d_1=1$, $d_2=0$, and $d_3=1$. We suppose that the noise sequence $\{w_k\}$ is iid with Gaussian distribution $\mathcal{N}(0,0.025)$. Denote
\begin{align*}
&\theta=[-a_1~(b_1d_1)~(b_1d_2)~(b_1d_3)]^T\\
&\varphi_k=[y_k~u_k~u^2_k~u^3_k]^T.
\end{align*}
Then the Hammerstein system can be formulated as
\begin{align}
y_{k+1}=\theta^T\varphi_k+w_{k+1}.\label{Eq86}
\end{align}

Let the reference signals $\{y^*_k\}$ be given by
$$
y^*_k\!=\!
\begin{cases}
+1,k\in\big[1000l+1,\cdots,1000l+500\big]\\
-1,k\in\big[1000l+501,\cdots,1000l+1000\big]
\end{cases}\!\!,l\geq 0.
$$

\begin{figure}
\centering
\includegraphics[width=10cm,trim=100 0 80 0, clip]{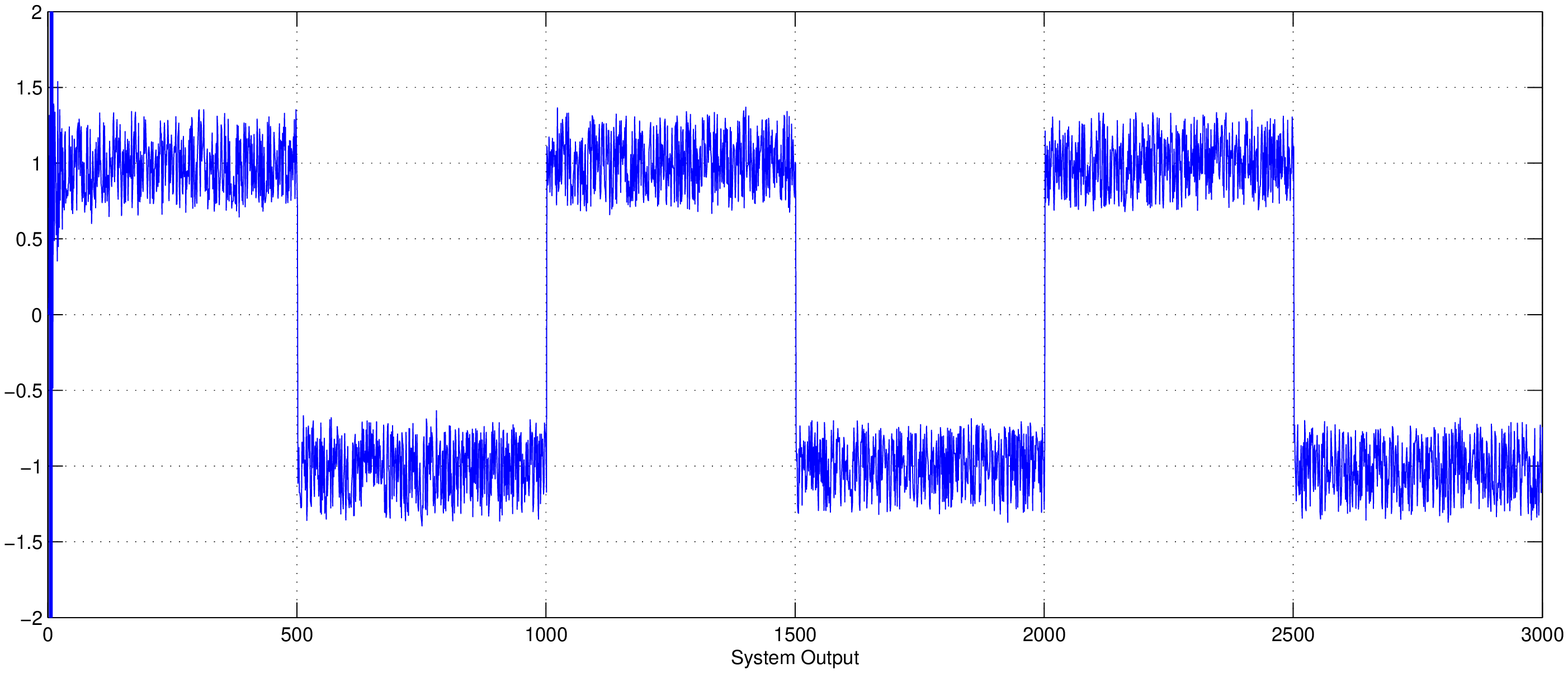}
\caption{System Output under Self-Tuning Regulation Control with Diminishing Excitation}\label{Fig3}
\centering
\includegraphics[width=10cm,trim=100 0 80 0, clip]{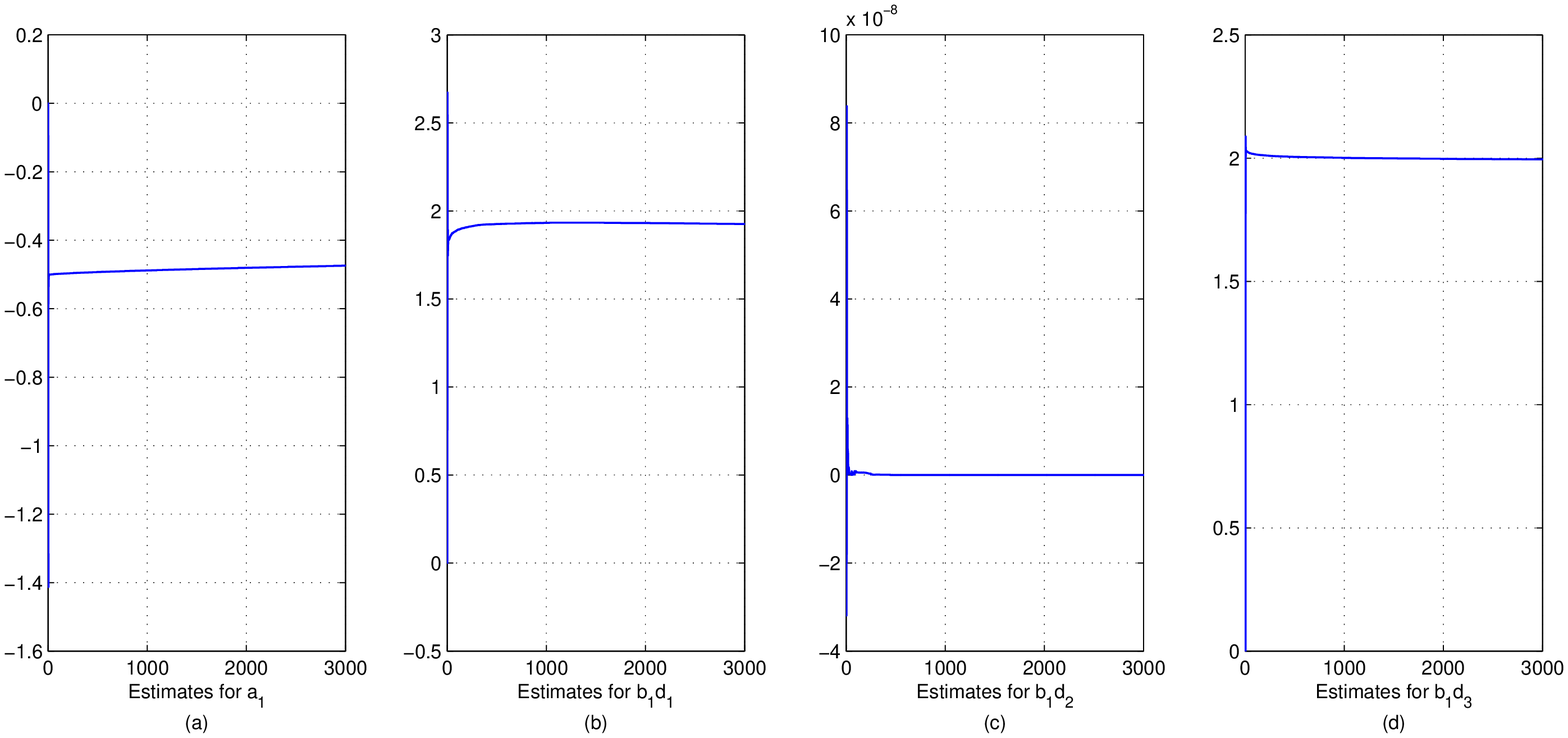}
\caption{Estimation Sequences $\{a_{1,N},(b_1d_1)_N,(b_1d_2)_N,(b_1d_3)_N\}_{N=1}^{3000}$}\label{Fig4}
\end{figure}

Denote the least squares estimates for $\theta$ by $\{\theta_k\}_{k\geq1}$. By noticing that $f(\cdot)$ is a third-order polynomial, the self-tuning regulation control with diminishing excitation is given by
\begin{align}
&u^0_k=\mathrm{RealSolution}\{u~|~y^*_{k+1}=-a_{1,k}y_k+(b_1d_1)_ku+(b_1d_2)_ku^2+(b_1d_3)_ku^3\},\label{Eq87}\\
&u_k=u^0_k+\frac{w'_k}{r_{k-1}^{\varepsilon/2}},\label{Eq88}
\end{align}
where $\mathrm{RealSolution}(\cdot)$ means the real solution of the third-order polynomial with minimal magnitude, $r_{k-1}=1+\sum\limits_{l=1}^{k-1}\|\varphi_l\|^2$, $\varepsilon=1/15$, and $\{w'_k\}$ is an iid sequence uniformly distributed over $[-0.1,0.1]$ and independent of $\{w_k\}$. Figure \ref{Fig3} shows the system outputs under the feedback control (\ref{Eq87})--(\ref{Eq88}) and Figure \ref{Fig4} shows the estimates $\{a_{1,N},(b_1d_1)_N,(b_1d_2)_N,(b_1d_3)_N\}_{N=1}^{3000}$ generated by (\ref{Eq5})--(\ref{Eq8}). Table \ref{tab5} compares the least squares estimates and the estimates generated from (\ref{Eq5})--(\ref{Eq8}) for $b_1d_2$. From Figure \ref{Fig4} and Table \ref{tab5} we see that, under the feedback control, we can still correctly identify the zero and nonzero elements in the unknown parameters by algorithms (\ref{Eq5})--(\ref{Eq8}).

\begin{table*}[htbp!]\scriptsize
\caption{Comparison between Least Squares Estimates and Estimates Generated from (\ref{Eq5})--(\ref{Eq8})}
\begin{center}
\begin{tabular}{|c|c|c|c|c|c|}
\hline & N=100 & N=200 & N=300 & N=400 & N=500 \\

\hline \begin{tabular}{c}estimates for $b_1d_2$\\ from (\ref{Eq5})--(\ref{Eq8})\end{tabular} & $8.1846\times 10^{-10}$ & $5.1160\times 10^{-10}$ & $1.0079\times 10^{-10}$ & $8.1142\times 10^{-11}$ & $8.5915\times 10^{-12}$ \\
\begin{tabular}{c}estimates for $b_1d_2$\\ by least squares\end{tabular} & $0.0311$ & $0.0304$ & $0.0293$ & $0.0284$ & $0.0270$\\
\hline
\end{tabular}
\end{center}
\label{tab5}
\end{table*}

\section{Concluding Remarks}

In this work, we introduce a sparse identification algorithm based on $L_2$ norm with $L_1$ regularization and establish both the set and parameter convergence of estimates for systems possibly operating in the feedback control framework. The condition in this work significantly extends the irrepresentable conditions required in literature on the same topic. For future research, it will be interesting to consider the asymptotical normality and convergence rate of the proposed identification algorithm.

\renewcommand{\baselinestretch}{1.5}

\end{document}